\documentclass[twocolumn,floatfix]{aastex63}

\usepackage{amssymb}
\usepackage{amsmath}
\usepackage{hyperref}
\usepackage{utfsym}

\usepackage{graphicx}
\begin{document}

\title{Direct Evidence for Stellar Initial Mass Function Variation in the Milky Way} 

\correspondingauthor{Charles Steinhardt}
\email{csteinhardt@missouri.edu}

\author[0000-0003-3780-6801]{Charles L. Steinhardt}
\affiliation{Department of Physics and Astronomy, University of Missouri, 701 S. College Ave., Columbia, MO 65203}

\author[0009-0000-0378-286X]{Carter Meyerhoff}
\affiliation{Department of Physics and Astronomy, University of Missouri, 701 S. College Ave., Columbia, MO 65203}

\author[0009-0000-6816-7198]{Alexander J. Luening}
\affiliation{University of Rochester, 500 Joseph C. Wilson Blvd., Rochester, NY 146275}
\affiliation{Department of Physics and Astronomy, University of Missouri, 701 S. College Ave., Columbia, MO 65203}

\begin{abstract}
Because direct measurements require resolved stellar populations including low-mass stars, determining the stellar initial mass function (IMF) has been a historically difficult problem even within our own Galaxy and impossible everywhere else.  As a result, even though it is predicted that the IMF should vary depending upon the properties of each individual star-forming molecular cloud, it is standard to assume a Universal IMF.  Using recent observations from {\em Gaia}, it is now possible to test for IMF variation using resolved stellar populations in open clusters and a parameterization that separates properties of the IMF from subsequent dynamical evolution.  Here, we show that the IMF is not Universal but instead varies across individual Galactic stellar populations, reflecting evolution in the average conditions of molecular clouds over cosmic time.  This evolution is consistent with the predictions of a simple astrophysical model in which the IMF is environmentally-dependent and the Milky Way reflects typical galactic behavior in recent cosmic history.  Thus, observational evidence now agrees with long-standing theoretical and numerical predictions.  

\end{abstract}


\section{Introduction}

The light emitted by a typical galaxy is generally dominated by rare, massive stars that comprise only a small fraction of its stellar mass and between $10^{-3}$ and $10^{-6}$ of the full halo mass.  The remainder of an unresolved galaxy and its dark matter halo are essentially invisible and must be inferred from these high-mass stars.  Estimating the mass-to-light ratio requires several strong assumptions, of which perhaps the most significant is the stellar initial mass function (IMF), or mass distribution of stars at formation.

Measuring the IMF even locally has been one of the historically difficult problems in astronomy, with several different observed IMFs still in common use \citep{Salpeter1955,Kroupa2001,Chabrier2003}.  Whichever Galactic IMF is chosen, standard techniques assume it to be Universal, applying to all galaxies under all conditions \citep{Brammer2008,Speagle2014}.  Previous observations have been consistent with a nearly-Universal IMF, although they have only been possible for very local stellar populations.

However, the fragmentation process producing individual stars from a collapsing molecular cloud should be sensitive to the speed of sound $c_s$ in that cloud \citep{LyndenBell1976,Larson1985} (and other effects; \citealt{Hopkins2012}).  Thus, changing, e.g., the gas temperature should produce a family of different IMFs \citep{Jermyn2018,Steinhardt2020b}.  Fitting templates derived using this family of IMFs \citep{Sneppen2022} to photometric catalogs of unresolved galaxies indeed found that most galaxies have higher sound speeds than in Galactic clouds \citep{Steinhardt2022a,Steinhardt2022b}, and therefore should also be expected to have bottom-lighter IMFs.  The resulting difference in galaxy properties is most significant at high redshift \citep{Steinhardt2023c}, where conditions in star-forming regions are least similar to Galactic clouds today.

All of these conclusions depend on the single-parameter functional form used for the family of potential IMFs.  If the wrong IMF is adopted, all of the inferred parameters, including even basic galaxy properties, will be significantly biased.  This sensitivity has historically led to skepticism regarding IMF variation, particularly given the risk of overfitting to photometric observations with relatively few bands.  This work addresses that concern by identifying clear evidence of IMF variation in open clusters within the Milky Way, where the underlying stellar populations can be observed directly. 

\S~\ref{sec:imf} demonstrates that the choice of a \citet{Kroupa2001} IMF allows a separation of properties of the initial mass function from subsequent dynamical evolution in open clusters.  The \citet{Hunt2023} {\em Gaia} catalog and quality cuts are described in \S~\ref{sec:catalog}.  In \S~\ref{sec:models}, these clusters are then used to show that the IMF indeed does vary between clusters, consistent with the predictions of simple physical models and consistent with the evolution in the average conditions of molecular clouds over time.  The implications of these results for extragalactic analysis are discussed in \S~\ref{sec:discussion}.

\section{Initial Mass Function Parameterization}
\label{sec:imf}

The initial mass function corresponds to the stellar mass distribution of an age-zero stellar population, and thus of an open cluster at formation.  Here, a \citet{Kroupa2001} IMF, 
\begin{equation}
\label{eq:kroupaimf}
    \xi(m) = \frac{dN}{dm} \propto \begin{cases}
    m^{-0.3}, & m < 0.08 M_\odot \\
    m^{-1.3}, & 0.08 M_\odot < m < 0.50 M_\odot \\
    m^{-2.3}, & m > 0.50 M_\odot \\
    \end{cases}
\end{equation}
is assumed for open clusters formed today.  For the clusters in this work, the {\em Gaia} detection limit is above the low-mass breakpoint, so only the intermediate-mass and high-mass regimes can be observed.  Similar astrophysical results would be produced using other functional forms (e.g., a \citet{Chabrier2003} IMF).  However, the Kroupa IMF allows an unambiguous interpretation of the results below, whereas the effects of a variable IMF and dynamical evolution are degenerate for a Chabrier IMF (see \S~\ref{subsec:kroupa}).  

A significant complication is that what can be measured in clusters are present-day stellar mass functions, whereas the quantity of interest is the initial mass function at formation.  For older clusters, the current stellar mass function (CMF) should differ from a Kroupa mass function due to three effects:
\begin{itemize}
    \item The most massive stars have the shortest lifetimes, and will thus no longer be present.  This is a well studied effect which has long been used to find the ages of clusters \citep{Sandage1953}.  
    \item Dynamical disruption preferentially strips low-mass stars over time, creating a progressively bottom-lighter SMF as the cluster ages.  Although the effect has been studied for decades \citep{Spitzer1958}, it is more complex than stellar aging.  Here, the results of an N-body simulation showing that cluster disruption is dominated by tidal stripping \citep{Gieles2008,PortegiesZwart2010} are used to model the effect as a function of cluster age.
    \item Galactic conditions have changed over time, and older clusters generally formed in hotter molecular clouds at lower metallicity.  A corresponding increase in the speed of sound produces a bottom-lighter IMF for those clusters \citep{LyndenBell1976,Larson1985,Jermyn2018,Steinhardt2020b}.  The results of a simulation modeling the effects of changing environment on the IMF are consistent with this analytical prediction \citep{Guszejnov2022}.
\end{itemize}

\begin{figure*}[htb!]
    \centering
\includegraphics[width=0.90\textwidth]{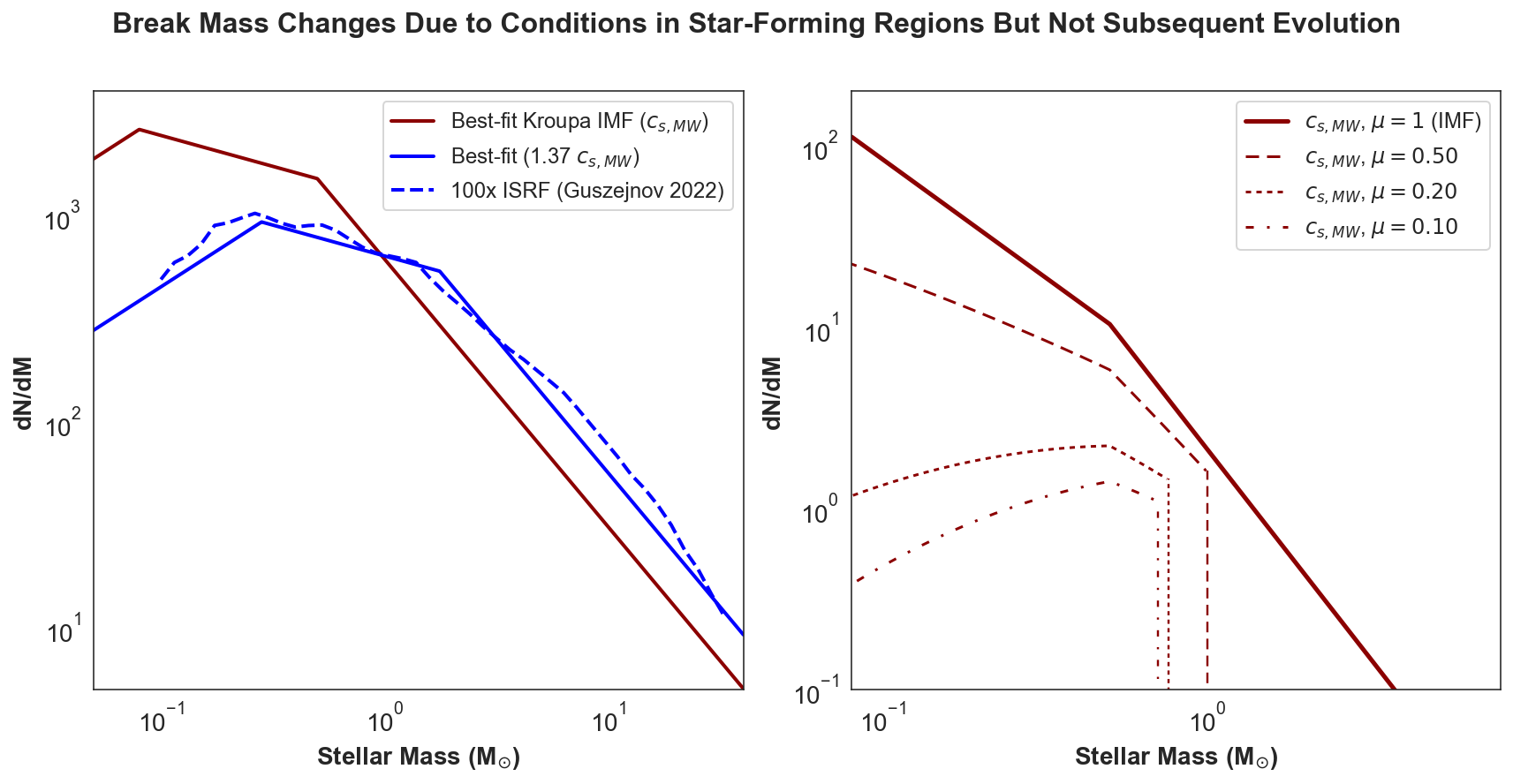}
    \caption{(Left) Theory predicts (blue, solid) that an increase in the sound speed will increase the break masses compared with a Kroupa IMF (red) but not the slopes.  A simulation increasing the input stellar radiation field by a factor of 100 \citep{Guszejnov2022} shows a similar increase in break masses, although the temperatures and sound speed within the simulated cloud exhibit a complex profile rather than a single value. (Right) Evolution of the stellar mass function over time (dashed) for an open cluster with a Kroupa IMF (solid), showing the effects of tidal stripping and stellar evolution, based on the semi-analytical approximations in \citet{Lamers2013} (see also \citealt{Gieles2008,PortegiesZwart2010,Lamers2013}).  The time required for individual clusters to reach remaining mass fractions of $\mu = 0.5$, 0.2, and 0.1 will vary depending upon cluster parameters.  The slopes and high-mass cutoff evolve over time, but the break mass is not affected. Thus, an observed change in break mass must be due to the IMF rather than subsequent evolution.}
    \label{fig:model}
\end{figure*}

These three astrophysical mechanisms have easily distinguishable effects on the mass functions of open clusters (Fig. \ref{fig:model}).  The deaths of massive stars are evident in the high-mass cutoff, also used to find the age of each cluster.  Dynamical effects increase both the intermediate-mass and high-mass slope, but do not change the breakpoint.  Finally, a change in $c_s$ does not affect the slopes but will shift the breakpoint, with older clusters predicted to exhibit breaks at higher mass than younger clusters.  Therefore, measuring the break mass and determining whether it is identical across different open clusters provides a direct test of IMF variation.

\subsection{Importance of Kroupa IMF Parameterization for Evaluating Dynamical Effects\\}
\label{subsec:kroupa}

It is common to choose any of several different functional forms for the IMF, including the Kroupa-like form used here, a Chabrier-like or log-normal form, and even a Salpeter-like simple power law.  It is also generally assumed that as long as only one of these is used consistently throughout the analysis, the key conclusions will be robust to a change in which functional form is selected.  Surprisingly, that is not true here: the analysis in this paper requires the use of a Kroupa IMF in order to clearly distinguish between IMF variation and dynamical evolution.  

The Kroupa IMF used here was initially selected for two other reasons.  First, the astrophysical derivation provided by Kroupa \citep{Kroupa2001} allows a more direct determination of the effects of changing environment on the IMF than empirically-derived mass functions.  Second, simulations line up well with a Kroupa-like functional form \citep{Guszejnov2022}.  

\begin{figure*}
    \centering
\includegraphics[width=0.9\textwidth]{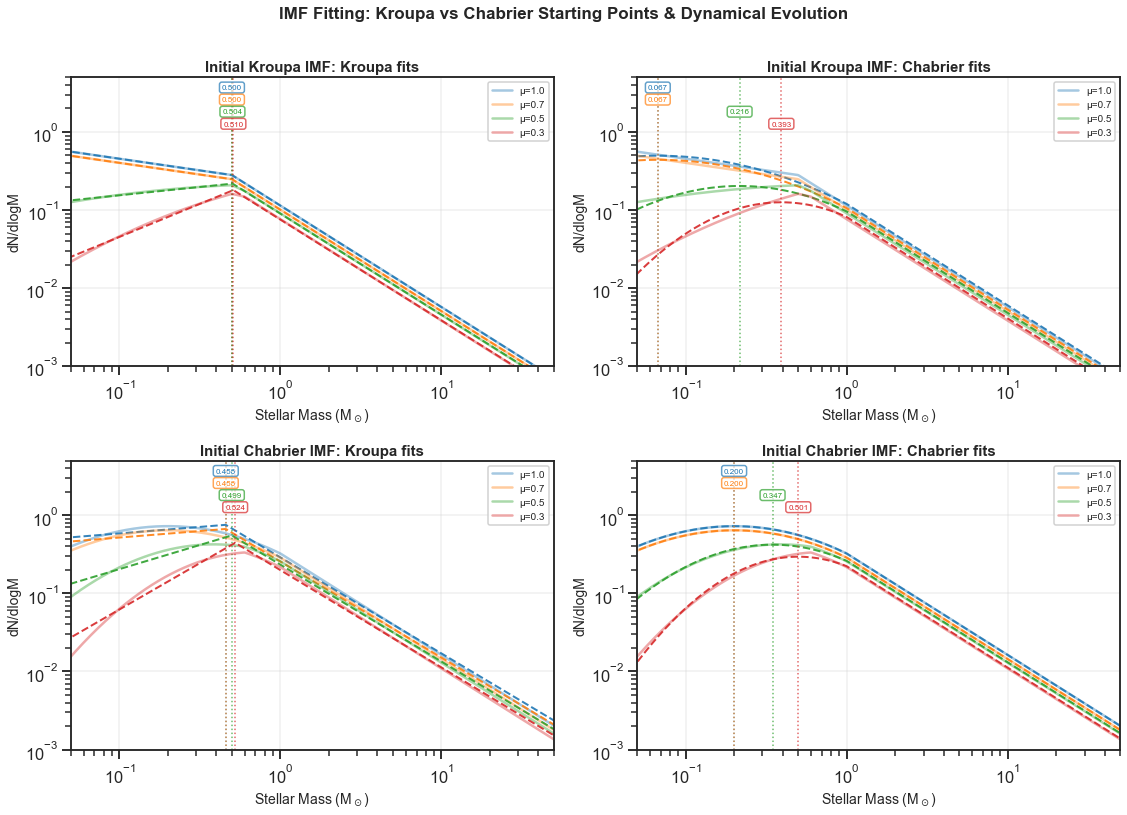}
    \caption{Variation in best-fit Kroupa (left) and Chabrier (right) mass functions as open cluster stellar populations are dynamically depleted according to the \citet{Lamers2013} prescription, for both a Kroupa (top) and Chabrier (bottom) initial mass function.  Depletion prior to mass segregation (yellow) primarily alters the normalization.  After mass segregation, further depletion increases the characteristic mass for the best-fit Chabrier mass function, but the break mass for a Kroupa mass function is nearly invariant.  This occurs regardless of whether the true underlying IMF is Kroupa-like or Chabrier-like.  Thus, the choice of a Kroupa mass function breaks the degeneracy between dynamical evolution and IMF variation which exists for a Chabrier fit.}
    \label{fig:kroupavschabrier}
\end{figure*}

However, there is a third property which is essential for determining properties of the IMF from the open cluster CMF.  The cumulative effects of dynamical evolution lead to a mass-dependent depletion of cluster stars that is well approximated by a power law.  Multiplying that power law with a Kroupa IMF consisting of multiple power laws with a break mass will produce a new Kroupa-like IMF, with different power law slopes but an identical break mass (Fig. \ref{fig:kroupavschabrier}).  Thus, any shift in the break mass can only come from IMF variation, not subsequent dynamical evolution.

By contrast, a power-law modification of a log-normal distribution, or of a Chabrier IMF, does not produce a new, Chabrier-like function.  The resulting CMF is no longer well described by a Chabrier-like parameterization, leading to systematic mismatches between the true evolved distribution and the model.  When such an evolved CMF is nonetheless fit with a Chabrier function, the inferred characteristic mass shifts even in the absence of any change in the underlying IMF (Fig. \ref{fig:kroupavschabrier}).  Thus, apparent evolution in the characteristic mass cannot be uniquely attributed to IMF variation, as it is degenerate with the effects of dynamical evolution.  

This degeneracy is particularly difficult to break for older clusters, which both are likely to have formed in different environments and simultaneously exhibit greater mass depletion simply due to their age, producing a natural correlation between star-forming environment and mass depletion.  Previous studies observing an increase in the characteristic mass for older clusters thus attributed it entirely due to mass depletion, which is correlated with the shift in characteristic mass.  However, a Kroupa-like IMF responds differently, since mass depletion only modifies the slopes, leaving the break mass unchanged.  Adopting a Kroupa functional form in this work allows variations in the break mass due to the IMF to be isolated from dynamical evolution, so that the interpretation is unambiguous.

\section{Cluster Catalog and Analytical Techniques}
\label{sec:catalog}

\subsection{Cluster Catalog and Sample Selection}

The open clusters for this paper and the methods used to identify them are described in detail in \citet{Hunt2023} and \citet{Hunt2024}. This catalogue contains a total of 7167 clusters detected in \emph{Gaia} DR3, of which 3530 are highly reliable open clusters. All of these clusters have measured parameters required for this study such as distance, age, extinction, or mass.  Although the Hunt \& Reffert catalogue provides estimated cluster ages, in some cases more precise determinations are available. Ages were therefore taken, in priority order, from \citet{Bossini2019}, \citet{Dias2021}, and \citet{CantatGaudin2020}, with Hunt \& Reffert ages used for the remaining clusters. Clusters in the Bossini et al. catalog without stated age uncertainties were assumed to have uncertainties typical of well-measured clusters of a similar age.

Additional quality cuts have been made for this paper to ensure results were only drawn from high-quality astrometric data and well-measured clusters, including both rare, high-mass stars and low-mass stars below the break:
\begin{itemize}
    \item {\bf Type `o' or `m'}: The Hunt \& Reffert catalogue includes a cluster type flag, and only open clusters are included in this analysis.
    \item {\bf Bins outside break}: In order to robustly determine the break mass, only clusters with at least three mass bins on each side of the best-fit break were included.  A jackknifing test indicated that this was the minimum number of bins required to robustly determine the break mass.
    \item {\bf Statistical Significance}: Only well-determined clusters with high statistical significance were chosen to reduce probability of randomness.
    \item {\bf CMD}: Only clusters with a high-quality CMD flag, indicating high confidence that they consist of a single stellar population with few outliers, are included.     
    \item {\bf Extinction}: To prevent dust from reducing quality of cluster measurements, only clusters with low extinction values were included.  This should also mitigate any effects of differential extinction on mass estimates.
    \item {\bf Slope Cuts}:  To ensure that a handful of points with correlated errors were not interpreted as a break in the mass function, only fits with a Kroupa-like high-mass slope and an intermediate-mass slope clearly distinct from the high-mass slope were included.
    \item {\bf Break Mass Error}: To accurately determine the change in break mass between clusters, objects with high break mass error were not included.  Although there is no correlation between distance and cluster properties in the full sample, a correlation is induced by this cut, since at large distances only clusters with a high break mass will contain enough low-mass stars to robustly determine the break.  As a result, there is a correlation between break mass and distance, but this correlation is not causal.  A generalized additive model (GAM) is used to correct for this induced bias.  As should be expected, after correction the statistical significance of the age-break mass relationship increases.
    \item {\bf Age Error}: To ensure accurate modeling of cluster parameters with age, only clusters with well measured ages were included.
\end{itemize}

Although the final sample includes only 110 of the 3530 high-quality clusters available, this is not primarily driven by quality cuts.  Only 417 of the 3530 clusters have sufficient mass completeness to detect low-mass stars below the break.  Of those, relatively few clusters also contain a sufficient number of the rare, high-mass stars above the break.

\subsection{Cluster Mass Functions\\}

The mass functions used here are based on those derived in the \citet{Hunt2024} catalog. In their work, accurate per-cluster mass functions were calculated by correcting raw stellar mass functions for several different effects. First, individual stellar masses (assuming single stars) were derived by interpolating PARSEC v1.2s isochrones \citep{BressanMarigo_2012} against \emph{Gaia} G-band magnitudes\footnote{Recent results suggest that these isochrones may induce a bias at low mass \citep{WangFan2025}.  If so, the specific break mass for each cluster might be biased, but this error would be systematic.  Thus, the quantitative location and evolution of the break mass might change, but the qualitative results should not.}.  Next, their mass functions were corrected for three selection effects: those of \emph{Gaia} data itself \citep{Cantat-GaudinFouesneau_2023}, of the selection of \emph{Gaia} data that they used \citep{Castro-GinardBrown_2023}, and of member stars missed by their adopted clustering algorithm. Neglecting these selection effects can otherwise strongly bias the slopes of cluster mass functions and cause severe underestimation of slope uncertainties, particularly at the faint end \citep{Hunt2024}. Finally, they also corrected their mass functions for unresolved binary stars by assuming each cluster contains an unresolved binary population described by relations in \citet{MoeDiStefano_2017}, which causes an additional small upwards correction to all mass bins on the order of $\approx$20\% \citep{Hunt2024}. 

\begin{figure}
    \centering
\includegraphics[width=0.45\textwidth]{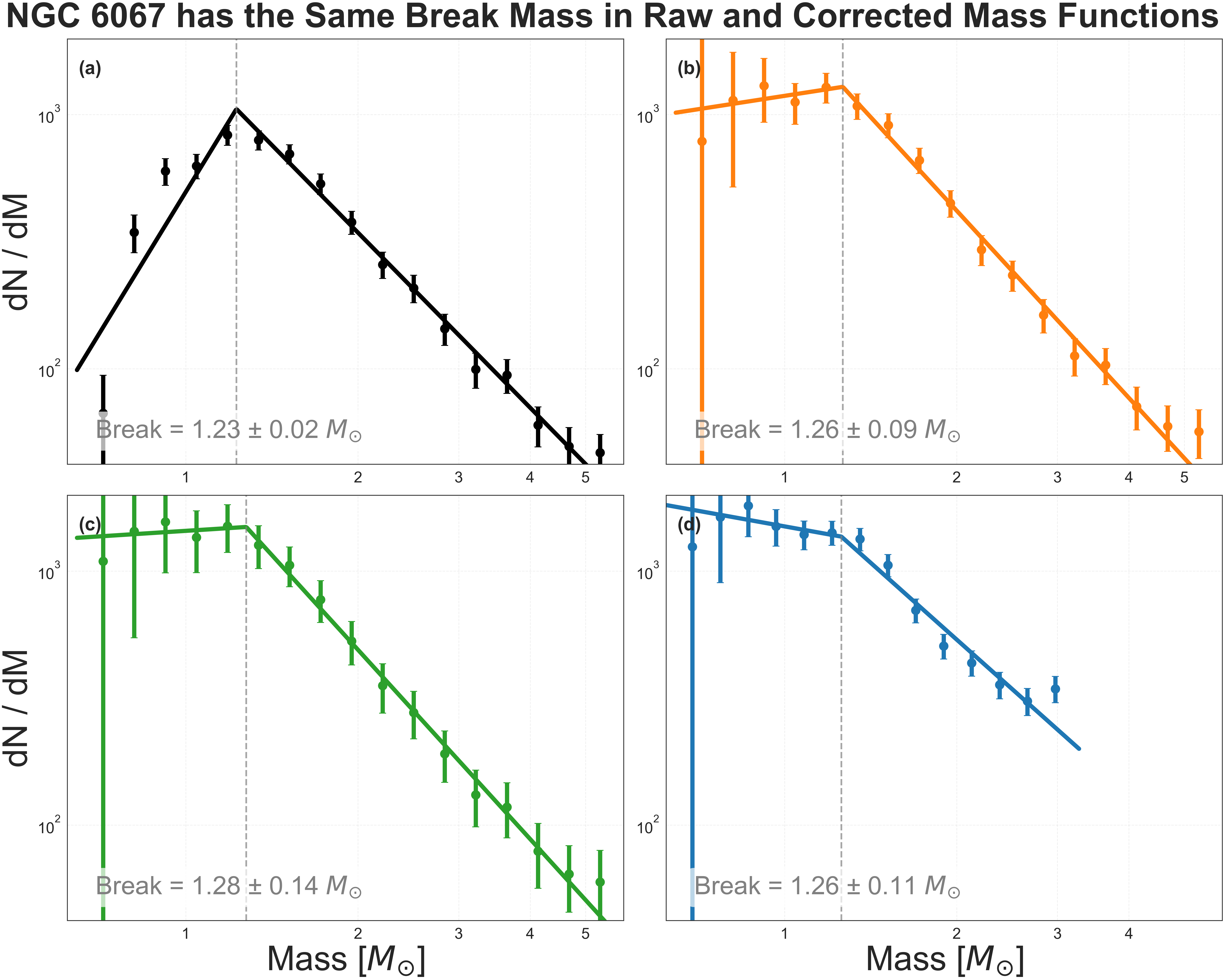}
    \caption{Mass functions of NGC 6067 at various stages of the analysis all yield similar break points, demonstrating that the existence and location of the break mass are features of the data rather than induced by analytical techniques.  (a) Raw stellar counts already exhibit a break mass at $1.23 \pm 0.02,M_\odot$.  (b) Applying a correction for {\em Gaia} selection changes the slopes but not the break mass.  (c) Applying a further correction for unresolved binaries yields a break mass of $1.28 \pm 0.14,M_\odot$ for the mass function used in this work.  (d) If an incorrect cluster age of 4 Gyr is used instead of the correct $1.25$ Gyr, the break mass remains unchanged.}
    \label{fig:cmfsteps}
\end{figure}

Although these corrections are essential for obtaining reliable cluster stellar mass functions, they do not change the fundamental result. In Fig.\ref{fig:cmfsteps}, this is illustrated using NGC 6067 as an example. The raw counts already reveal a clear break mass at $1.23 \pm 0.02,M_\odot$. Applying corrections for \emph{Gaia} selection effects and unresolved binaries significantly changes both the high-mass and intermediate-mass slope but shifts the break mass only slightly, to $1.28 \pm 0.14,M_\odot$.  Indeed, the key results in this paper about break mass evolution can be seen even from raw counts, although these selection corrections are required for rigorous analysis.  Break mass evolution has also been found in previous studies \citep{deMarchi2010,Bastian2010,Offner2014,Moraux2016,Jiang2024}, which used different methodology and completeness corrections.

The analysis also relies on first measuring the cluster age in order to choose a specific isochrone.  However, even when an incorrect cluster age of is assumed (4 Gyr instead of the correct $1.25$ Gyr in Fig.\ref{fig:cmfsteps}), the inferred break mass remains unchanged.  Thus, the observed variation in break masses between different clusters, and their correlation with age, are robust features of the data rather than products of the analysis techniques.

\section{Initial Mass Function Variation in Open Clusters}
\label{sec:models}

\begin{figure}
    \centering
\includegraphics[width=0.45\textwidth]{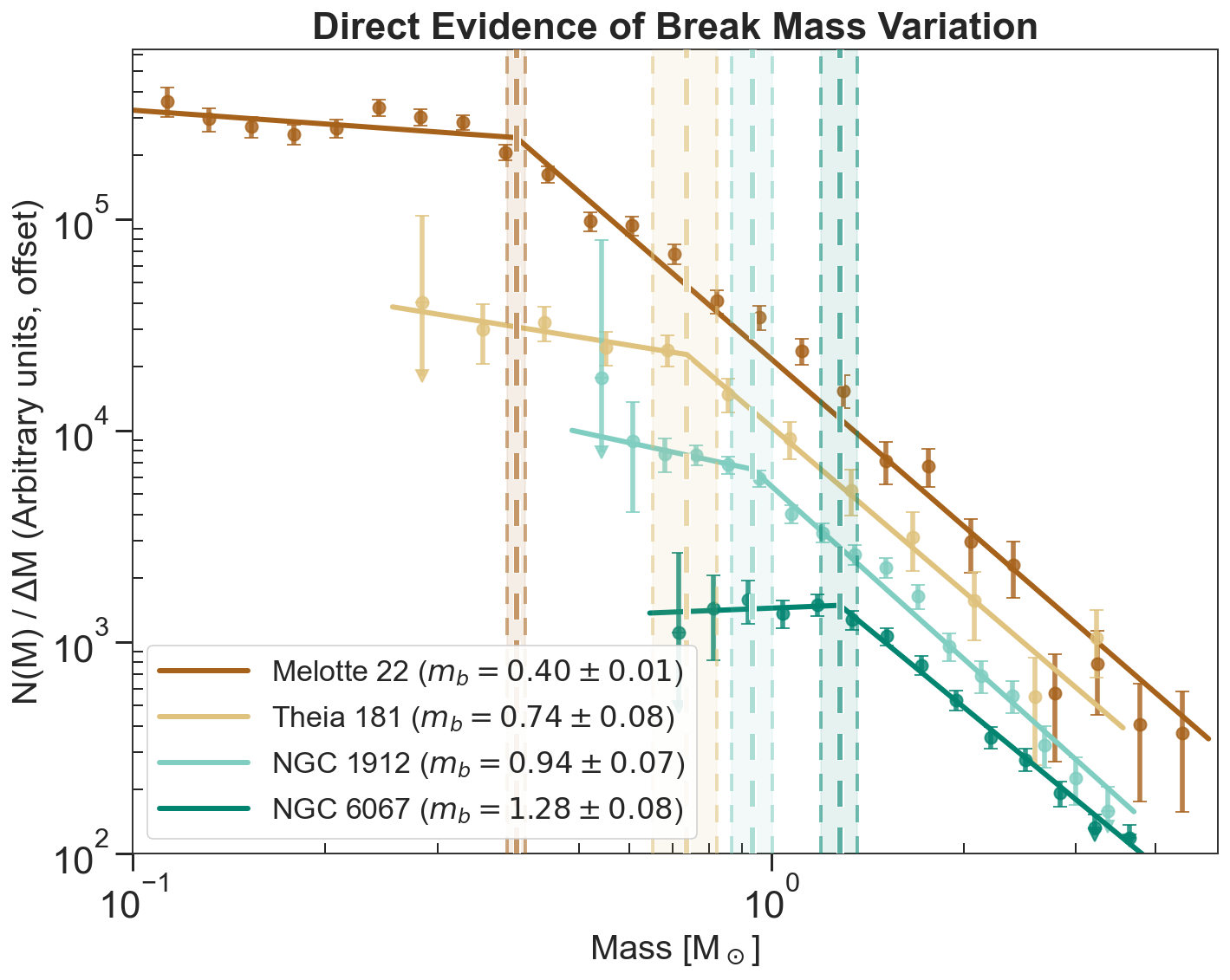}
    \caption{Observed stellar mass functions for four different open clusters from the \citet{Hunt2023} catalog.  The break masses are different for each cluster.  This cannot be due to subsequent stellar evolution or dynamical disruption, and thus must be caused by variation in the IMF.}
    \label{fig:fits}
\end{figure}

Clusters exhibit a range of different break masses (Fig. \ref{fig:fits}), as predicted assuming they formed in molecular clouds with a variety of sound speeds.  This observed variation in break mass is consistent with several previous studies, which also found break mass variation in the current mass functions of open clusters \citep{deMarchi2010,Bastian2010,Offner2014,Moraux2016,Baumgardt2023,Jiang2024}.  These studies typically concluded that the shift in break mass was not due to a variable IMF, but rather due to dynamical effects that first segregate the cluster by stellar mass and then preferentially strip low-mass stars, so that the CMF will be bottom-lighter than the IMF.  Jiang et al. found a correlation between greater mass segregation and a higher break mass \citep{Jiang2024}.

However, both simulations and semi-analytical approximations find that these dynamical effects are scale-free \citep{Baumgardt2003,Lamers2013} (Fig. \ref{fig:model}).  Thus, dynamical effects will produce shallower intermediate-mass slopes in clusters with greater mass segregation, as previously found, but {\em cannot change the break mass}.  Further, although some clusters do appear to undergo surpisingly rapid dynamical evolution \citep{DellaCroce2024}, typical dynamical timescales are longer than the $10^7 - 10^9$ yr ages of the clusters used here.  So, any age dependence must instead reflect differences in the conditions in star-forming regions at the time of cluster formation. 

\begin{figure*}
    \centering
\includegraphics[width=\textwidth]{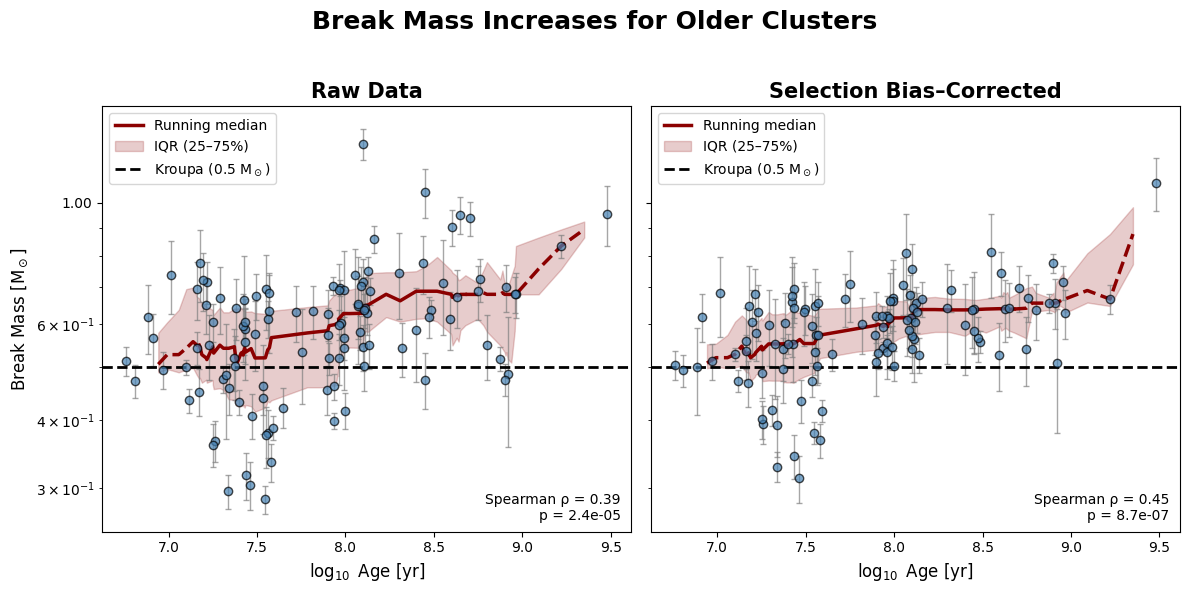}
    \caption{(left) The break mass observed in Gaia clusters (points, with running median and interquartile range in red, dashed for few clusters in window) increases with cluster age.  (right) As expected, selection biases artificially steepened the slope and added scatter.  After correcting for the bias, the slope decreases but the correlation strengthens, revealing the true underlying break mass-age relation.  The trend is consistent with the sound speed evolution of a typical galaxy \citep{Steinhardt2023b}, which will produce an increasing break mass for older clusters.  A Kroupa IMF (dashed), defined to have a break mass of $0.5 M_\odot$, is shown for comparison.}  
    \label{fig:breakmass}
\end{figure*}

Indeed, time evolution of this sort has already been predicted.  Galaxies are broadly observed to follow evolutionary tracks in which gas temperatures decline and metallicities increase over time \citep{Steinhardt2023b}.  On average, older clusters are therefore expected to form at lower effective sound speeds ($c_s$).  Although the high-mass slope (Spearman $\rho = 0.055$, $p = 0.57$) and intermediate-mass slope (Spearman $R\rho = -0.153$, $p = 0.06$) show no statistically significant correlation with cluster age in the current sample, the break mass is indeed strongly correlated with cluster age ($\rho = 0.391$, $p = 2.4 \times 10^{-5}$)\footnote{Uncertainties in both variables, as in this case, are known to reduce the measured $\rho$, so $2.4 \times 10^{-5}$ is an upper bound on the $p$ value.  Even extensive simulations failed to produce any null realizations, so that the true significance of the correlation could not be estimated.}, increasing for older clusters (Fig.~\ref{fig:breakmass}, left).  This implies a long-term evolution in the average star-forming conditions within the Milky Way consistent with trends in a typical galaxy.  

However, the observed age–break mass relation is subject to distance-dependent selection effects.  The accessible survey volume increases with distance while the detection threshold limits the observable parameter space to systems with sufficiently high break masses (preferentially not just older clusters but also scattered above the intrinsic relation) at large distances.  This produces a truncation bias that distorts the apparent slope and scatter of the relation.  To correct for this bias, the joint dependence of break mass on age and distance is modeled using a generalized additive model (GAM; \citealt{Hastie1986}).  The residual age–break mass relation after accounting for the distance-dependent component (Fig.~\ref{fig:breakmass}, right) has a lower slope, but stronger correlation ($\rho = 0.449$, $p = 8.7 \times 10^{-7}$) as would be expected after correcting for this bias.

However, the rate of decline in temperature can vary substantially between galaxies.  Moreover, the Milky Way might have an atypical star-formation history.  A galaxy remaining on the observed star-forming main sequence with a present-day star formation rate of $1\,M_\odot\,{\rm yr^{-1}}$ and stellar mass of $10^{11}\,M_\odot$ would typically have formed stars much more rapidly a few Gyr ago, yielding a steeper decline in $T_{\rm IMF}$.  The Milky Way \citep{delAlcazar2025}, however, appears to have maintained a comparatively constant SFR over that period, suggesting a more gradual decline in $T_{\rm IMF}$ than average.  Thus, although the trend is qualitatively consistent with the expected decline in gas temperatures over time, significantly improved modeling would be needed to test the break mass evolution quantitatively for the Milky Way.

In addition to the broader trend, clusters at fixed age span a substantial range of break masses, inconsistent with a single value.  This, too, is expected from previous observations, which show that galaxies are not monolithic but rather exhibit a wide range of conditions in different regions at the same epoch \citep{Sanchez2012,Bundy2015}.  These variations in the physical conditions of star-forming molecular clouds will produce a corresponding range of sounds speeds, and thus break masses.

The same effect might help to reconcile previous studies using local stellar populations to measuring the IMF with the results in this work.  Several studies using stellar populations within 20-100 pc have found IMFs consistent with a standard Kroupa IMF and a breakpoint around $0.4-0.5 M_\odot$ \citep{Sollima2019,Kirkpatrick2024,Wang2025}.  When multiple regions with different IMFs are averaged together, the resulting aggregate population tends to resemble a single IMF of the same shape \citep{Hunt2024}, particularly in small volumes that sample fewer environments.  By contrast, {\em Gaia} observations combined with spectroscopic ages and metallicities allow stars to be subdivided into subsamples in the much larger volume from 100-300 pc, revealing IMF variation \citep{Li2023}.

\section{Discussion}
\label{sec:discussion}

{\em Gaia} observations of the stellar mass functions of open clusters show clear evolution in the characteristic intermediate-to-high-mass break with cluster age, while both the intermediate- and high-mass slopes remain largely unchanged. This evolution is observed across the full subset of clusters with sufficient completeness and membership to constrain the break, and the results are incompatible with a fixed, universal IMF.  The relationship between break mass and cluster age is statistically robust ($p = 5.2 \times 10^{-6}$) and indicates a time-dependent IMF whose properties evolve in a specific and measurable way.

The observed evolution in break mass with cluster age is consistent with a physically motivated model relating fragmentation within a molecular cloud to the speed of sound rather than relying on population synthesis or empirical templates.  This model was previously found to be consistent with hydrodynamical simulations and now is directly supported by observations of resolved stellar populations.  Thus, there is now a consensus between theory, simulation, and observations on the evolving nature of stellar initial mass functions. 

The implications of an evolving IMF extend well beyond the Milky Way.  Because the light from a galaxy is dominated by only a tiny fraction of its stellar mass, it is necessary to assume an IMF to infer stellar masses, star formation rates, and other properties.  Most current techniques rely on population synthesis models that assume a fixed IMF.  

If the IMF varies with physical conditions, these properties will not only be incorrect but strongly biased, potentially altering conclusions about star formation, mass assembly, and feedback across cosmic time.  The effect will be most significant for JWST observations of ultra-high-redshift galaxies, whose physical conditions differ most from the current environment of the Milky Way.  In particular, both stellar masses and star formation rates are likely significantly overestimated, so that seemingly extreme measurements \citep{Labbe2023} should become easier to reconcile with standard astrophysical and cosmological models \citep{Steinhardt2023}.

In summary, these observations provide strong evidence against the long-standing assumption of a universal IMF.  Indeed, there is now a growing consensus between theory, simulation, and observations on the evolving nature of stellar initial mass functions. The agreement between observations of open clusters and astrophysical models in which the IMF depends upon environment suggests that IMF variation is a necessary component of feedback in models of stellar and galaxy evolution. A universal IMF should no longer be a default assumption, but rather only the local limit of a broader astrophysical model for star formation.

\section{Acknowledgements}

The authors would like to thank Emily Hunt for help with the Hunt \& Reffert cluster catalog, providing intermediate data products, and for her advice on \emph{Gaia}-related issues.  They would also like to thank Holger Baumgardt and Lachlan Hobart for assistance in ensuring that the mass depletion due to dynamical effects shown in this work matched existing simulations.  Finally, the authors would like to thank Yicheng Guo, Stella Offner, Albert Sneppen, John Weaver, Christopher Wikle, and Haojing Yan for helpful comments.  Carter Meyerhoff was funded by a Mizzou Forward Undergraduate Research Training Grant and the Summer Research Exposure Program.


This work has made use of data from the European Space Agency (ESA) mission {\it Gaia} (\url{https://www.cosmos.esa.int/gaia}), processed by the {\it Gaia} Data Processing and Analysis Consortium (DPAC,
\url{https://www.cosmos.esa.int/web/gaia/dpac/consortium}). Funding for the DPAC has been provided by national institutions, in particular the institutions participating in the {\it Gaia} Multilateral Agreement.

\bibliographystyle{aasjournal}
\bibliography{refs.bib} 

\appendix

\section{Robustness of Break Mass Variation}

\begin{table}[!h]
    \centering
    \begin{tabular}{|c|c|c|c|c|c|}\hline
         &  & \multicolumn{2}{c|}{Raw Correlation} & \multicolumn{2}{c|}{Bias-Corrected} \\\cline{3-6}
         & Clusters & Spearman $\rho$& $p$ value & Spearman $\rho$& $p$ value \\\hline
         All cuts & 110& 0.391& $2.4 \times 10^{-5}$& 0.449& $8.7 \times 10^{-7}$\\\hline
         Type 'o' or 'm' & 110& 0.391& $2.4 \times 10^{-5}$& 0.449& $8.7 \times 10^{-7}$\\\hline
         Bins outside break $>$ 2 & 113& 0.373& $4.7 \times 10^{-5}$& 0.401& $1.1 \times 10^{-5}$\\\hline
         Statistical Significance $>$ 15 & 139& 0.402& $9.3 \times 10^{-7}$& 0.435& $8.7 \times 10^{-8}$\\\hline
         Color Magnitude Diagram $>$ 0.75 & 112& 0.387& $2.5 \times 10^{-5}$& 0.449& $ 6.6 \times 10^{-7}$\\\hline
         Extinction $<$ 1.0 & 115& 0.377& $3.3 \times 10^{-5}$& 0.464& $1.8 \times 10^{-7}$\\\hline
         High mass slope quality & 168& 0.448& $1.3 \times 10^{-9}$& 0.401& $ 8.1 \times 10^{-8}$\\\hline
         Intermediate mass slope quality & 116& 0.406& $6.1 \times 10^{-6}$& 0.499& $1.2 \times 10^{-8}$\\\hline
         $m_{\mathrm{break}}$ error $<$ 0.15 & 116& 0.371& $4.2 \times 10^{-5}$& 0.417& $ 3.2 \times 10^{-6}$\\\hline
         Age error $<$ 0.3 & 120& 0.415& $2.4 \times 10^{-6}$& 0.474& $4.7 \times 10^{-8}$\\\hline
    \end{tabular}
    \caption{To test whether the correlation between break mass and cluster age is driven by any particular quality cut, the analysis was repeated while removing each cut in turn.  In every case, there is strong statistical evidence for positive correlation between break mass and age, both in the raw data and after correcting for selection bias.}
    \label{tab:cuts}
\end{table}

In order to restrict the sample to clusters with robustly-determined break masses, several substantial quality cuts were made.  As a result, it is essential to ensure that cuts did not bias the eventual conclusions.  Table \ref{tab:cuts} shows the effects of removing each individual cut on the sample size and break mass-age correlation. No single cut made a significant difference, and in every case there is strong statistical evidence for positive correlation between break mass and age ($p \lesssim 10^{-5}$ for the null hypothesis).  A Pearson $R$ produces comparable $p$ values, and a jackknife test finds an average $p$ value of $1 \times 10^{-6}$, compared to $9 \times 10^{-7}$ for the full sample.

\begin{figure}
    \centering
\includegraphics[width=0.9\textwidth]{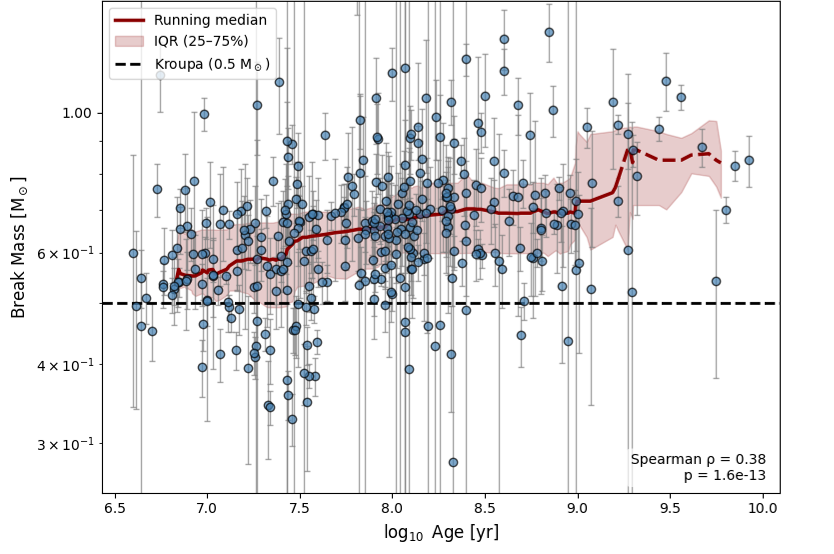}
    \caption{Looser quality cuts (see Table \ref{tab:relaxedcuts}) produce a similar result to the main sample shown in Fig. \ref{fig:breakmass}.  There is a correlation of comparable strength between age and break mass, with older clusters tending to have higher break masses.  Due to the much larger sample, the statistical significance of this correlation increases from $2 \times 10^{-5}$ to $10^{-13}$.  Thus, the conclusion that there is a trend consistent with the sound speed evolution of a typical galaxy is not an artifact of quality cuts, but instead reflects a genuine property of the underlying cluster population.}
    \label{fig:loosercuts}
\end{figure}

If every cut is relaxed substantially (Table \ref{tab:relaxedcuts}), the sample size expands to 358 of the 417 clusters.  The same analysis as in Fig. \ref{fig:breakmass} now produces a comparable correlation strength but with much higher statistical significance ($p = 10^{-13}$ compared with $2 \times 10^{-5}$ for the restricted sample; Fig. \ref{fig:loosercuts}) due to the larger sample size.  Thus, the quality cuts do not induce the correlation, but merely restrict the sample to the clusters for which the break mass can be measured most reliably.

\begin{table}[!h]
    \centering
    \begin{tabular}{|l|c|c|}\hline
          &Final Cuts& Looser Cuts\\\hline
          Type 'o' or 'm'&Cut made& Cut Made\\\hline
          Bins Outside Break&$>$ 2& $>$ 0\\\hline
          Statistical Significance&$>$ 15& $>$ 7.5\\\hline
          Color Magnitude Diagram&$>$ 0.75& $>$ 0.50\\\hline
          Extinction&$<$ 1.0& $<$ 1.25\\\hline
          High Mass slope quality&-2.85 $<$ x $<$ -1.9& -5.0 $<$ x $<$ 5.0\\\hline
          Intermediate Mass slope quality&-1.7 $<$ x $<$ 1.0& -5.0 $<$ x $<$ 5.0\\\hline
          $m_{\mathrm{break}}$ error&$<$ 0.15& $<$ 0.50\\\hline
          Age Error&$<$ 0.30& $<$ 0.50\\ \hline
    \end{tabular}
    \caption{Comparison of quality cuts for the main sample of 110 clusters (Final Cuts) and the expanded sample of 358 clusters (Looser Cuts). The looser cuts allow more clusters to pass each criterion, resulting in a larger sample of generally lower-quality break mass measurements.  The larger sample produces similar trends to the main sample, with higher statistical significance due to the larger sample size.}
    \label{tab:relaxedcuts}
\end{table}

\label{lastpage}
\end{document}